\documentclass[fleqn,10pt]{SelfArx} 

\definecolor{color1}{RGB}{0,0,90} 
\definecolor{color2}{RGB}{0,20,20} 

\newlength{\tocsep} 
\usepackage{lipsum} 
\usepackage{times}
\usepackage{courier}
\usepackage{multirow}
\usepackage{float}
\usepackage{mathptmx}
\usepackage{amsmath}
\usepackage{epstopdf}
\usepackage[scaled=.90]{helvet}
\usepackage{subfigure}
\usepackage{graphicx}

\JournalInfo{Learning in Repeated Games: Human Versus Machine} 
\Archive{} 

\PaperTitle{Learning in Repeated Games: Human Versus Machine} 

\Authors{Fatimah Ishowo-Oloko\textsuperscript{1}, Jacob Crandall\textsuperscript{1}, Manuel Cebrian\textsuperscript{2,3}, Sherief Abdallah\textsuperscript{4,5}, Iyad Rahwan\textsuperscript{1,5,*}} 
\affiliation{\textsuperscript{1}\textit{Department of Electrical Engineering and Computer Science, Masdar Institute of Science and Technology, Abu Dhabi, UAE}} 
\affiliation{\textsuperscript{2}\textit{Department of Computer Science and Engineering, University of California at San Diego, La Jolla, California, United States of America,}} 
\affiliation{\textsuperscript{3}\textit{National Information and Communications Technology Australia, Melbourne, Victoria, Australia}}
\affiliation{\textsuperscript{4}\textit{British University in Dubai, United Arab Emirates}}
\affiliation{\textsuperscript{5}\textit{School of Informatics, University of Edinburgh, Edinburgh, United Kingdom}}
\affiliation{\textsuperscript{*}\textbf{Corresponding author}: irahwan@acm.org} 

\newcommand{\keywordname}{Keywords} 

\Abstract{While Artificial Intelligence has successfully outperformed humans in complex combinatorial games (such as chess and checkers), humans have retained their supremacy in social interactions that require intuition and adaptation, such as cooperation and coordination games. Despite significant advances in learning algorithms, most algorithms adapt at times scales which are not relevant for interactions with humans, and therefore the advances in AI on this front have remained of a more theoretical nature. This has also hindered the experimental evaluation of how these algorithms perform against humans, as the length of experiments needed to evaluate them is beyond what humans are reasonably expected to endure (max 100 repetitions). This scenario is rapidly changing, as recent algorithms are able to converge to their functional regimes in shorter time-scales. Additionally, this shift opens up possibilities for experimental investigation: where do humans stand compared with these new algorithms? We evaluate humans experimentally against a representative element of these fast-converging algorithms. Our results indicate that the performance of at least one of these algorithms is comparable to, and even exceeds, the performance of people.}

\begin{document}

\flushbottom 

\maketitle 

\tableofcontents 

\thispagestyle{empty} 


\section{Introduction} 

Since its inception, the field of Artificial Intelligence (AI) has both fascinated and frightened
human society. After Turing argued that machines can demonstrate intelligence \cite{Turing:1950}, both the scientific literature and popular culture explored the many possible future ramifications of this phenomenon, some of which are not so favorable to humanity. Indeed, AI technologies have been developing at a staggering speed, building largely on the exponential growth in computer processing power \cite{moore:1965} and increased algorithmic sophistication. Moreover, AI technologies are becoming pervasive in our everyday lives, with applications ranging from internet search and medical diagnosis, to fraud detection, stock market trading, and online dating.

The AI community has been very successful in developing AI that outperforms humans in complex analytical tasks, such as chess \cite{Campbell:etal:2002}, checkers \cite{Schaeffer:etal:2007}, and jeopardy \cite{Ferrucci:etal:2010}.  However, humans still retain their supremacy in tasks that require intuitive judgment \cite{Kasparov:2010}.  One example of such tasks are repeated social interactions, which can be modeled as repeated games.  Many repeated games have multiple (often infinite; see the folk theorem \cite{GintisGameTheory}) equilibria that limits the usefulness of analytical reasoning for prescribing successful behavior against arbitrary associates.  In these games, successful agents must resort to intuitive judgment to quickly adapt to unknown associates.

Humans are often able to arrive at cooperative solutions in these scenarios \cite{PassionWithinReason,Rilling:etal:2002}. However, human capacity to find such desirable solutions appears to rely on many psychological mechanisms that are not yet fully understood, though our understanding of how humans find such solutions is growing \cite{fehr2004social,rand2013human}.  In particular, many key human (behavioral) strategies have been identified theoretically \cite{Nowak:2006} and verified through a growing body of experimental evidence \cite{Dal:2011,Fudenberg:etal:2012}.

Despite this knowledge, AI algorithms for repeated interactions have been far less successful at finding cooperative solutions than humans for various reasons. Many algorithms produce myopic strategies that fail to learn profitable cooperative strategies in repeated games. They also adapt too slowly for realistic interaction, often requiring thousands of interactions to learn good behavior even in the simplest games. Some competitive algorithms have been extremely successful in special cases (e.g., Tit-for-tat in the Prisoners' Dilemma  \cite{Axelrod:1984} and algorithms for large zero-sum games such as Poker \cite{PolarisPokerChampion}), but finding successful algorithms for repeated general-sum games played against arbitrary associates has proven difficult.  This suggests that humans are ``still safe'' from machines in tasks that require social cooperation and coordination.

However, some recently-developed AI algorithms for repeated games now adapt on time scales applicable to human interaction.  Thus, for the first time, we can compare the performance of these algorithms with humans, both in interaction with humans and against other artificial agents.  In this paper, we give experimental evidence that human supremacy over machines in social games is no longer a given.  To show this, we performed a user study in which 58 participants were paired with each other and with AI algorithms in a variety of repeated general-sum games.  Our results show that a recently developed AI algorithm was, overall, more successful in these games than humans.  These results suggest that AI capabilities are catching up to humans in repeated interactions with (initially) unknown associates.

These results have implications for the ongoing debate about the risks of increasing automation, especially in environments in which distinguishing humans and machines is non-trivial.

\section{Algorithms for Repeated Games}
Algorithms for decision making in repeated games have been studied extensively over the last several decades.  One popular approach is to compute and play a desirable equilibrium strategy, which is chosen to meet a property of the game or the assumed tendencies of one's associate \cite{LittmanRepeatedNash,FolkEgal,Johansonetal2012}.   When assumptions are correct, this approach can be quite successful.  However, the presence of multiple (often infinite) equilibria in repeated general-sum games means that these assumptions are often not met, which sometimes results in the failure of the algorithm to produce effective strategies.

\begin{figure}
\begin{center}
	\includegraphics[width=3.4in]{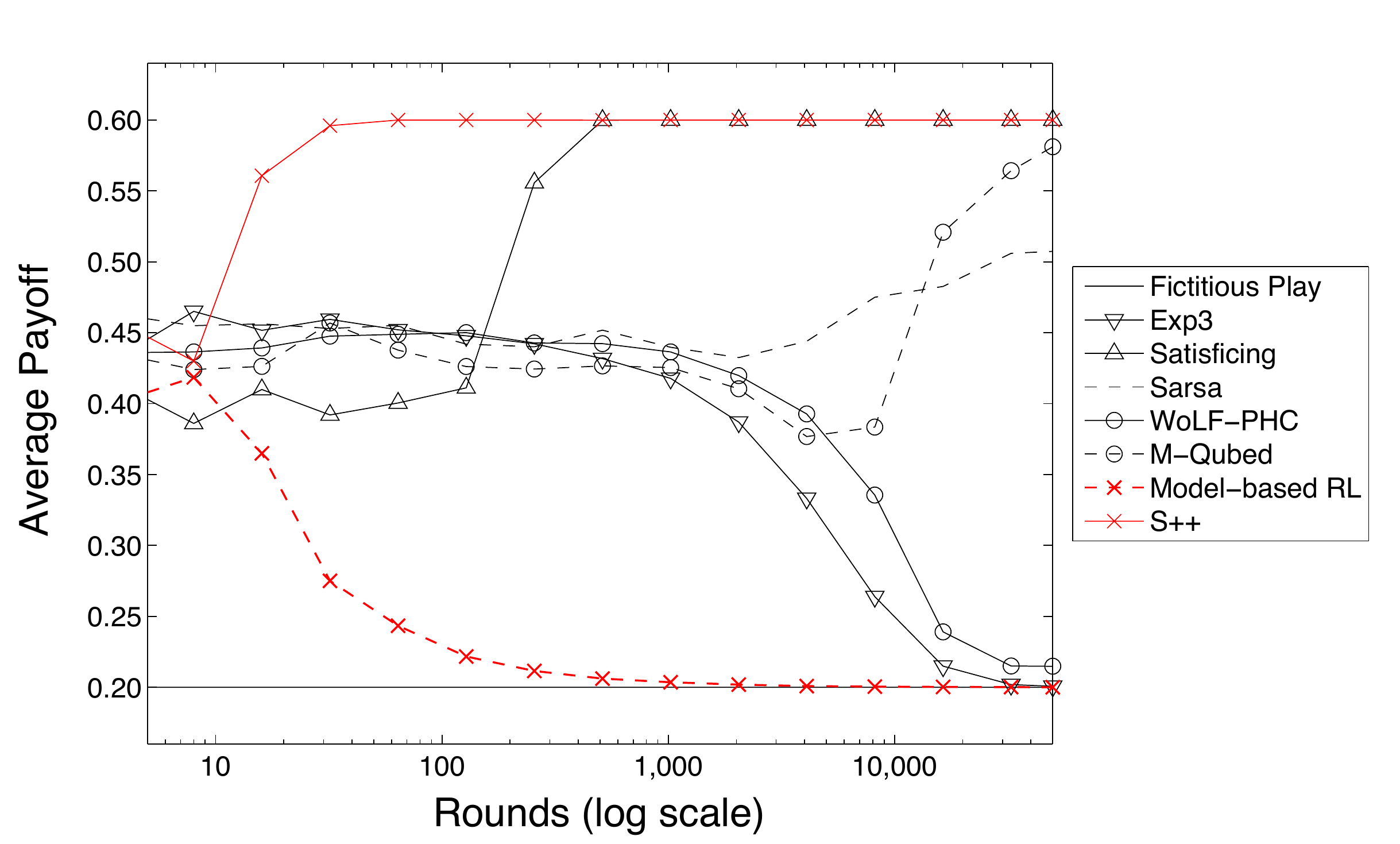}
	\caption{Average per-round payoffs over time of eight AI algorithms in self play in a prisoner's dilemma (Table~\ref{fig:gameMatrix}).   Results are the average payoffs of both players over 50 trials.  Only S++ converges to mutual cooperation (0.60) within 100 rounds, while Fictitious Play and Model-based RL converge to mutual defection (0.20) in less than 100 rounds.  The other algorithms require much more time to learn.}
\label{fig:speedtest}
\end{center}
\end{figure}

An alternative approach is to learn a desirable strategy in real time through experience. Influential learning algorithms developed for such environments use \emph{reinforcement learning} (RL) \cite{watkinsphd,MinimaxQ,friendorfoe,WoLF,CEQ,CrandallGoodrichMLJ2010}, \emph{opponent modeling} \cite{FictitiousPlay,GanzfriedSandholm2011}, \emph{aspiration learning} \cite{Karandikar,StimpsonIJCAI,Chasparis2010}, and \emph{expert algorithms}  \cite{UCB,exextrade,auer95gambling}.  However, until recently, no algorithm has been shown to learn effective strategies against arbitrary associates in general-sum games within time scales that are appropriate for interaction with humans (no more than 100 interactions).  Figure~\ref{fig:speedtest} gives a rough characterization of the performance of a number of these algorithms.  Several empirical studies  \cite{Frenchies,CrandallGoodrichMLJ2010} give a more complete picture of the performance of existing AI learning techniques for repeated games.  
 
Figure~\ref{fig:speedtest} shows the average per-round payoff over time of representative algorithms in self play in a repeated prisoner's dilemma (PD; see Table \ref{fig:gameMatrix}).  The figure shows that five of these algorithms (Exp3~\cite{auer95gambling}, Satisficing~\cite{StimpsonIJCAI}, Sarsa~\cite{Sarsa1994}, WoLF-PHC~\cite{WoLF}, and M-Qubed~\cite{CrandallGoodrichMLJ2010}) obtain average per-round payoffs of between 0.40 or 0.45 for well over the first 100 rounds, which represents exploratory (random) payoffs in this game --note that the x-axis is in log-scale.  Eventually, these algorithms typically learn in self play to either cooperate (which results in increased per-round payoffs approaching 0.60) or defect (which results in decreased per-round payoffs approaching 0.20).   While some of the algorithms are quite effective after hundreds or thousands of rounds of interaction, only three of these representative algorithms learn in time scales that are appropriate for interactions with a human: Fictitious play \cite{FictitiousPlay}, Model-based RL, and S++.  Since Fictitious Play is too myopic to learn profitable compromises in many games, we focus on Model-based RL and S++ in this paper.

Model-based RL is a well-known method for learning in many different scenarios.  We use this algorithm to represent traditional (older) AI techniques.  The particular version of this algorithm that we use in this paper encodes the state of the world as the last joint action played.  It then computes the state-transition model using the Fictitious-Play assessment (conditioned, again, on the previous joint action) from which it uses value iteration to compute a best response.  It then implements an $\varepsilon$-greedy strategy.  While it learns to defect in self play in the PD (Figure~\ref{fig:speedtest}), this algorithm quickly learns effectively against leader strategies \cite{LittmanLeaderAlgs} and other algorithms in some games.

S++ comes from a family of new (recently developed) expert algorithms that were shown to quickly learn near optimal behavior against a wide range of associates in repeated general-sum games \cite{Crandall_JAIR2014}.  These expert algorithms select from a set of leader and follower experts, each of which implements a particular equilibrium strategy.  A meta-algorithm that combines aspiration learning and reasoning over multiple opponent models trims the set of experts from which the expert algorithm selects experts in each time period.  Crandall showed that the resulting algorithms outperform other top-performing learning algorithms in a wide range of repeated games played against other AI algorithms.

Since Model-based RL and S++ both learn within time scales appropriate for interactions with humans, we use these algorithms to represent the progression of AI in repeated games.  Model-based RL represents traditional (older) AI algorithms, while S++ represents more recently developed AI algorithms.  In the remainder of this paper, we compare the performance of these algorithms with that of humans in various repeated games.  We are interested in investigating the ability of these algorithms to interact with humans and other algorithms in real time, using the performance of humans as the benchmark of success.

\section{Experimental Setup}
To compare the abilities of AI algorithms to that of humans in repeated interactions, we turn to repeated two-player normal-form games and, in particular, games of conflicting interest.  These games capture the tension between cooperation and competition. Such games have been used extensively to identify the mechanisms that promote cooperation in human and animal groups \cite{Nowak:etal:2011}. Specific mechanisms have been identified using theory and simulation \cite{Nowak:2006}, building on well-established models of population dynamics \cite{Nowak:2006bk}. Many such mechanisms have also been validated by experiments, in which human participants play different kinds of games in controlled laboratory settings \cite{Fehr:etal:2002,Rand:etal:2009,Rand:etal:2011}.

In two-player normal-form games, each player has a set of actions available to it.  Let $A_i$ denote the set of actions available to player $i$.  In each round of the game, each player $i \in \{1,2\}$ selects an action $a_i \in A_{i}$.  The resulting joint action $(a_1,a_2)$ produces a payoff vector $(R_1,R_2)$, which denotes the payoffs to players 1 and 2, respectively, in that round.  In all of the games we consider, the payoffs of the game are constrained to the range $[0,1]$.  The goal of each player is to maximize its payoff over the course of the repeated game.

In this study, we considered the four normal-form games depicted in Table~\ref{fig:gameMatrix}.  Each of these games represents a different common scenario in which cooperation and compromise is difficult to achieve against unknown associates.  Furthermore, each game has an infinite number of Nash equilibria (NE) of the repeated game, given that the game repeats after each round with high enough probability.  The first game is the Prisoners' Dilemma (PD). In the PD, defection (d) is the dominant strategy for both players. However, this joint action is Pareto dominated by the solution $(c,c)$ which leads to a payoff of $0.6$ for both players.  Both players can increase their payoffs by convincing the other agent to cooperate. The dilemma or challenge arises from the difficulty of convincing a rational player that it is better to play a dominated strategy. The pair $(c,d)$ means the column player gets the temptation reward of $1.0$ for defecting against its partner, while the row player gets the lowest payoff of $0$. Similarly, $(d,c)$ means the column player gets the lowest payoff of $0$ compared to the defecting partner's payoff of $1.0$.

Chicken, also known as the hawk-dove game in evolutionary game theory \cite{Smith:etal:1973}, models a situation where agents are in equilibrium when they choose opposing actions (i.e. one player chooses action $a$ while the other chooses action $b$ and vice versa). However, the agents differ in which equilibrium they prefer. Alternatively, the agents could both receive a high payoff if they both play $a$ which can also be made an equilibrium in repeated games given threats. Again, this represents a challenge for players on how to communicate their willingness to play a mutually beneficial solution through their actions.

Shapley's game, proposed by Lloyd Shapley \cite{Shapley:1964}, is a variation on rock, paper, scissors. This game is of interest because it has been shown that most learning algorithms do not converge when they play it \cite{FictitiousPlay}. The game does not have a pure one-shot NE, but rather a unique one-shot mixed strategy NE in which all agents play randomly. This NE yields a payoff of $1/3$ to each agent. However, in repeated interaction, both agents can do better by alternating between getting  a payoff of 0 and 1, which results in a Pareto-optimal outcome in which both players get an average payoff of $0.5$.

Chaos was designed by the authors such that the `right' action is not immediately obvious. Like the PD, the game has three Pareto-optimal solutions and a single one-shot NE solution in which both players play $a$. However, the game is less structured than the PD and has a larger strategy space, which seemingly would make cooperation and compromise more difficult to learn against arbitrary associates.

\begin{table}[t]
	\centering
	\subfigure[Prisoners' Dilemma (PD)]{
       		\label{tab:prisoners}
		{\small
          	\begin{tabular}{ccc} \hline
			& {\bf c} & {\bf d} \\ \hline
			{\bf c} & 0.60, 0.60 & 0.00, 1.00 \\
			{\bf d} & 1.00, 0.00 & 0.20, 0.20 \\ \hline
		\end{tabular}}}
	\subfigure[Chicken]{
       		\label{tab:chicken}
		{\small
          	\begin{tabular}{@{\hspace{.1cm}}c@{\hspace{.3cm}}c@{\hspace{.3cm}}c} \hline
			& {\bf a} & {\bf b} \\ \hline
			{\bf a} & 0.84, 0.84 & 0.33, 1.00 \\
			{\bf b} & 1.00, 0.33 & 0.00, 0.00 \\ \hline
		\end{tabular}}}
	\subfigure[Shapley's Game]{
       		\label{tab:shapleys}
          	{\small
		\begin{tabular}{@{\hspace{.1cm}}c@{\hspace{.25cm}}c@{\hspace{.25cm}}c@{\hspace{.25cm}}c} \hline
			& {\bf a} & {\bf b} & {\bf c} \\ \hline
			{\bf a} & 0, 0 & 1, 0 & 0, 1 \\
			{\bf b} & 0, 1 & 0, 0 & 1, 0 \\
			{\bf c} & 1, 0 & 0, 1 & 0, 0 \\\hline
		  \end{tabular}}}
	\subfigure[Chaos Game]{
       		\label{tab:chaos}
          	{\small
		\begin{tabular}{@{\hspace{.1cm}}c@{\hspace{.25cm}}c@{\hspace{.25cm}}c@{\hspace{.25cm}}c} \hline
			& {\bf a} & {\bf b} & {\bf c} \\ \hline
			{\bf a} & 0.46, 0.67 & 0.24, 0.06 & 1.00, 0.00 \\
			{\bf b} & 0.00, 0.37 & 0.37, 0.07 & 0.01, 0.53 \\
			{\bf c} & 0.14, 0.69 & 0.20, 1.00 & 0.71, 0.90 \\\hline
		  \end{tabular}}}
		
	\caption{Payoff matrices of four games considered in our study.  In each cell, the row player's payoff is listed first, followed by the column player's payoff.}
	\label{fig:gameMatrix}
\end{table}

Each participant in the study played all four repeated games, each of which against four different (unknown) associates.  The length of each game was unknown to the participants in advance to avoid end-game effects; each game was played for between 50 and 57 rounds (thus, we report only on results over the first 50 rounds).  The four different associates with which the participants were paired were another human participant, Model-based RL, S++, and a tit-for-tat style agent,\footnote{For clarity of exposition and since we only consider learning algorithms in this paper, we do not present results related to the tit-for-tat-style associate.  This algorithm performs very effectively in some scenarios and very poorly in others.  Overall, its performance was below that of S++ in the study.} which was included originally for control purposes.  The order of the games and associates experienced by the participants was varied throughout the study. Care was taken to ensure that the participants did not know who they were paired with at any time (human or artificial agent).  Participants joined the study in groups, typically of six people. The study protocol was approved by the university's ethics board.

\begin{figure*}
\centering
	\subfigure[Self play summary: All rounds]{\label{fig:selfBarChartAll}\includegraphics[width=3in]{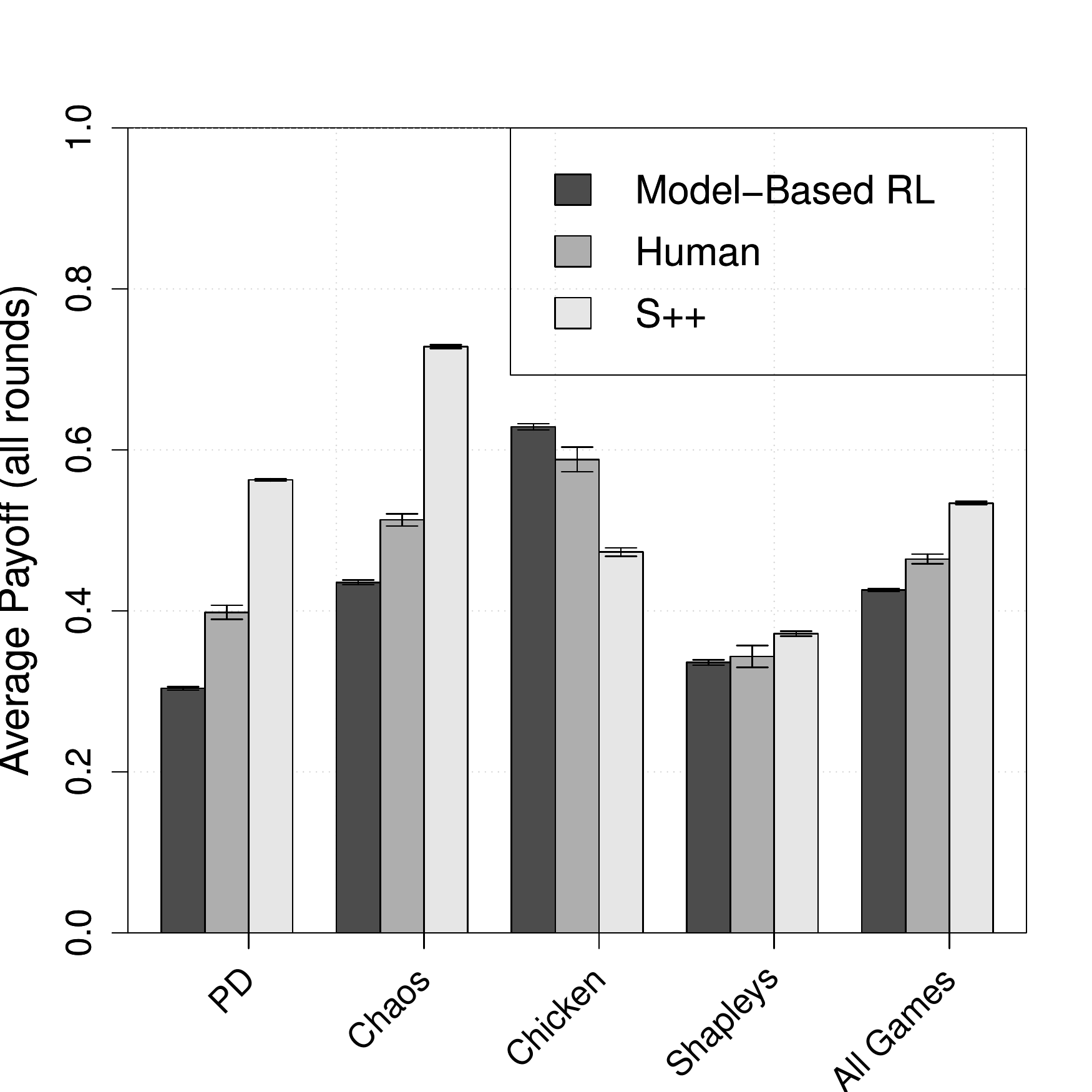}} ~~~~~~
	\subfigure[Self play summary: Last 10 rounds]{\label{fig:selfBarChartLast10}\includegraphics[width=3in]{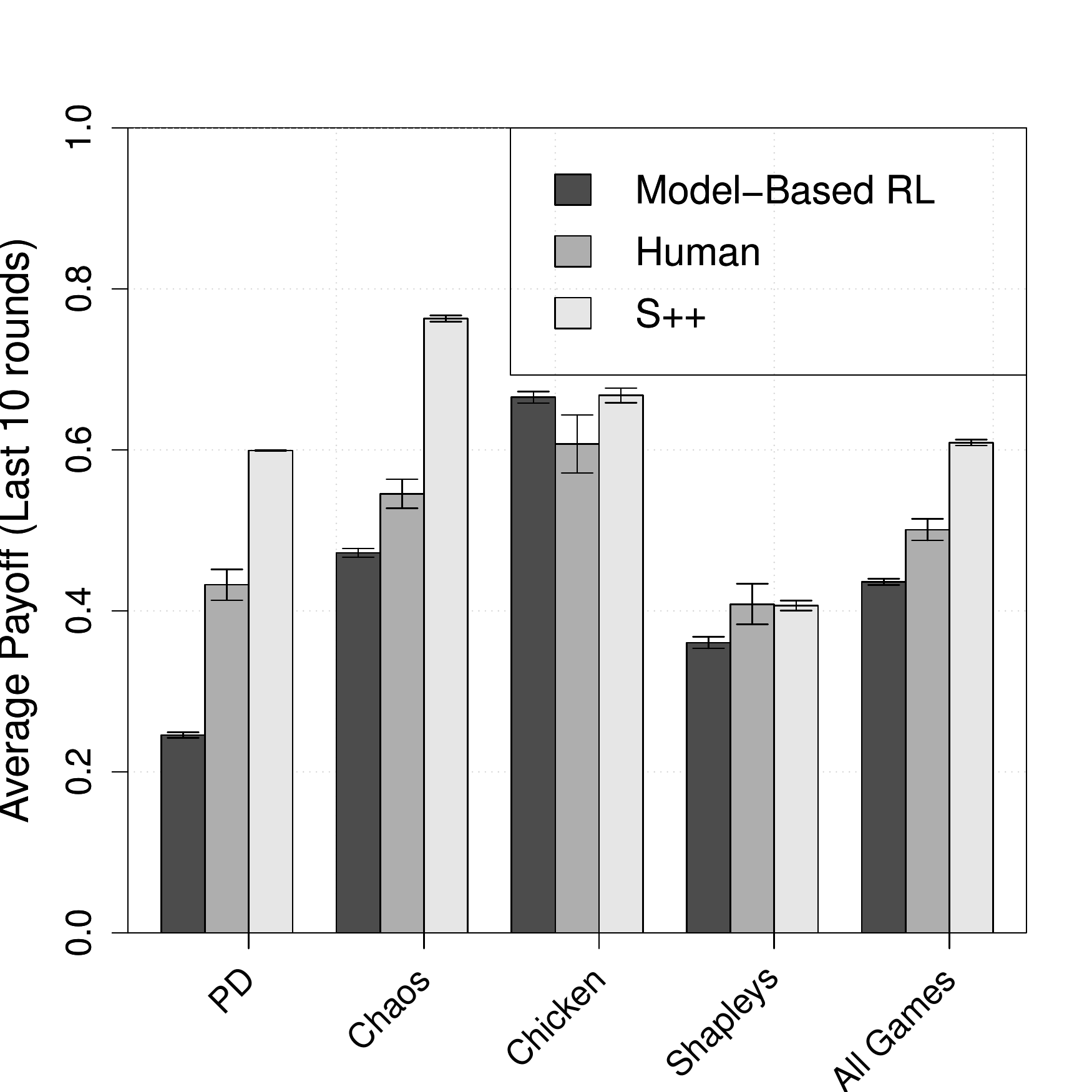}}\\
	\caption{Average payoffs of both players in self play in each game.  Error bars show the standard error of the mean.}
\label{fig:selfplay1}
\end{figure*}

The repeated games were played on a desktop computer.  All games were played with perfect information.  That is, participants were shown the payoff matrix on the computer screen, which showed the payoffs of both players in the game.  After selecting their action in each round of the game (by clicking the corresponding button), the participants were told the action chosen by their associate and the payoff received by both players in that round.  To give incentive to participants to do their best in the study, payment to the subjects was based on their performance.  Each participant received between \$5 and \$15, proportional to payoffs they received when playing the games. A total of 58 people from a university campus participated in the study.  The participants were almost exclusively graduate students and postdoctoral associates.



\section{Experimental Results}
We discuss the results of our study in two parts.  First, we consider how each agent (human or algorithm) performed in self play (i.e., scenarios in which both players are of the same type).  An algorithm's success in self play is critical to its evolutionary robustness~\cite{Axelrod:1984}.  Second, we evaluate the performance of the agents when they associate with each other.  These comparisons speak to the agents' abilities to interact successfully with arbitrary associates, which is necessary when players cannot pick their associates or the algorithms they use, such as with ad hoc teams \cite{StoneAdHoc}.

\subsection{Self Play}
\begin{figure*}
\centering
	\subfigure[Self play: Prisoner's dilemma]{\label{fig:pdSelfplay}\includegraphics[width=2.25in]{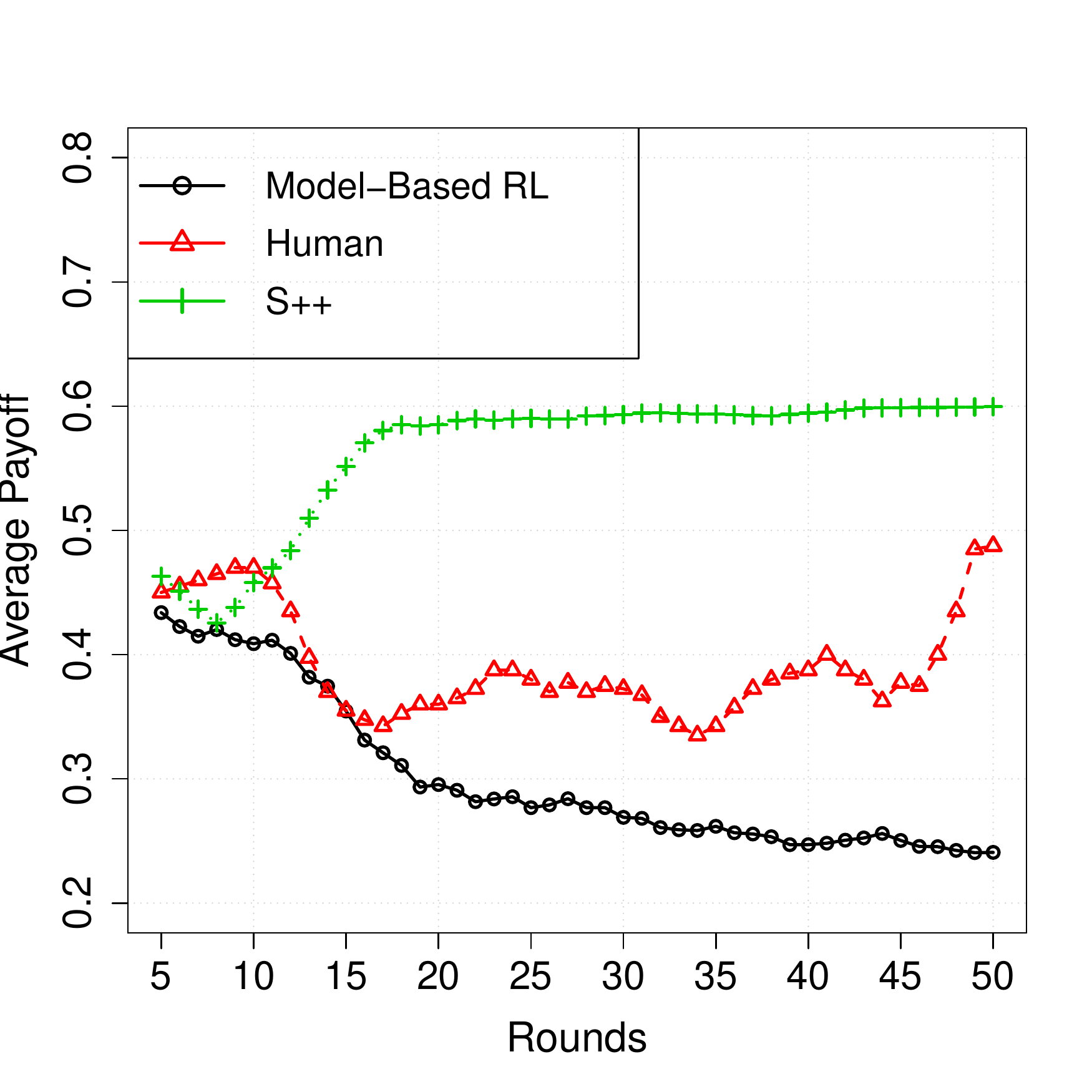}}
	\subfigure[Self play: Chaos]{\label{fig:chaosSelf}\includegraphics[width=2.25in]{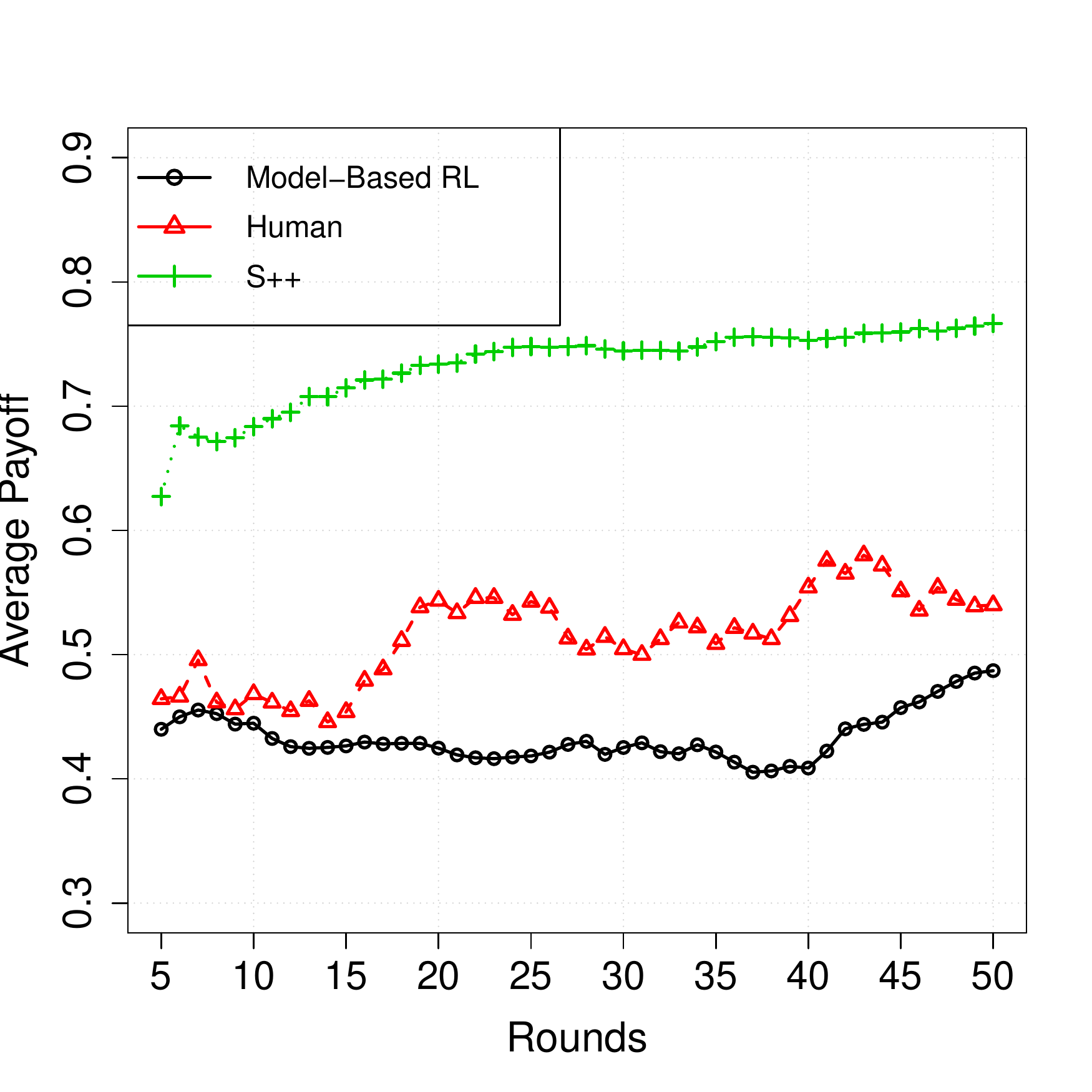}}
	\subfigure[Self play: Chicken]{\label{fig:chickenSelf}\includegraphics[width=2.25in]{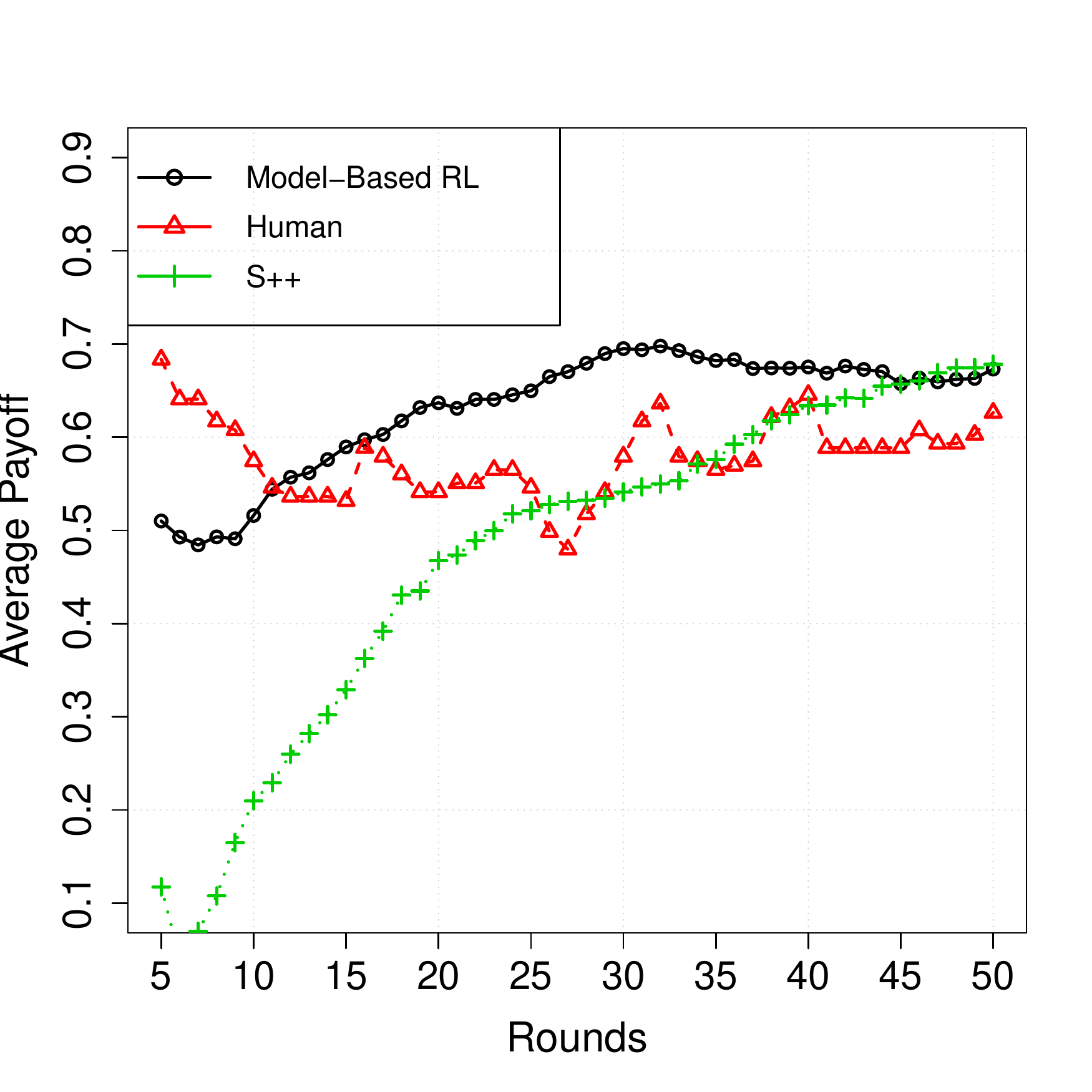}}\\ \vspace{-.15in}
	\subfigure[Self play: Shapley's game]{\label{fig:shapleysSelf}\includegraphics[width=2.25in]{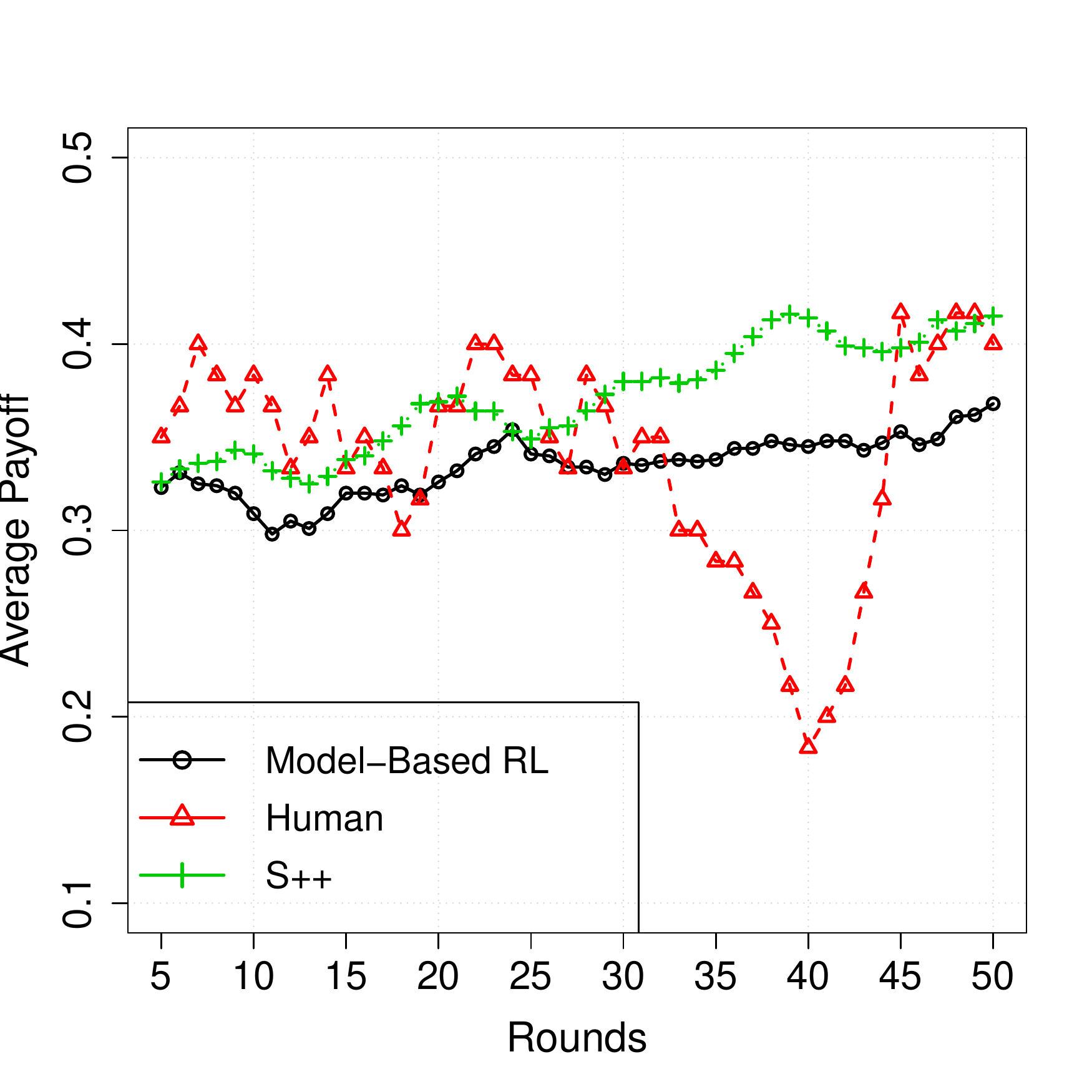}}
	\subfigure[Self play: All games]{\label{fig:allSelf}\includegraphics[width=2.25in]{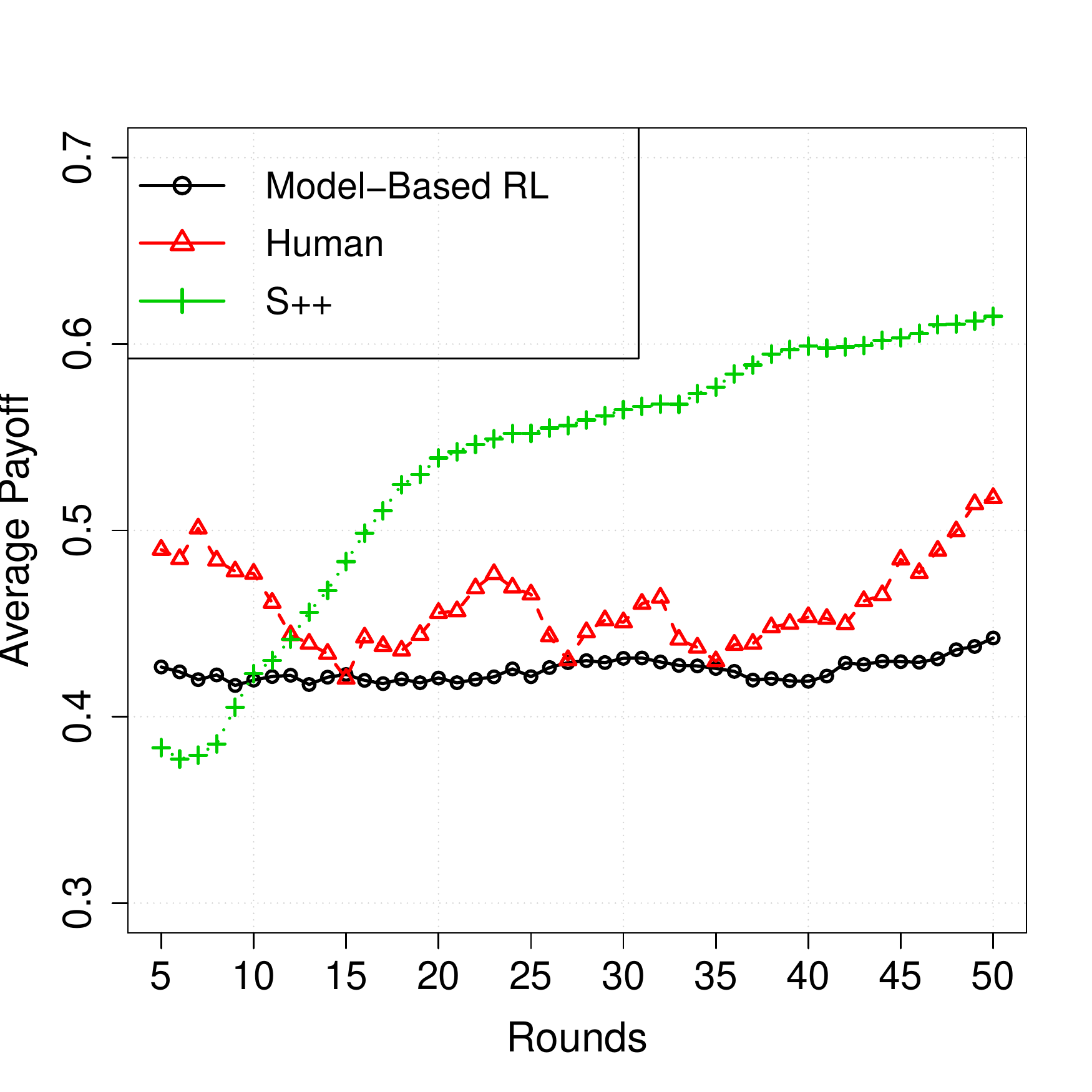}}
\caption{Average payoffs of both players over time in self play in each game.}
\label{fig:selfplay2}
\end{figure*}

The payoffs obtained by humans, S++, and Model-based RL in self play are summarized in Figures~\ref{fig:selfplay1} and~\ref{fig:selfplay2}.  Figure~\ref{fig:selfplay1} shows a general trend across the four repeated games used in the study.  While both S++ and humans outperformed model-based RL overall in self play, S++ obtained higher average payoffs than did humans.

\begin{figure*}
\begin{center}
	\subfigure[vs.~Humans: All rounds]{\label{fig:hBarChartAll}\includegraphics[width=2.25in]{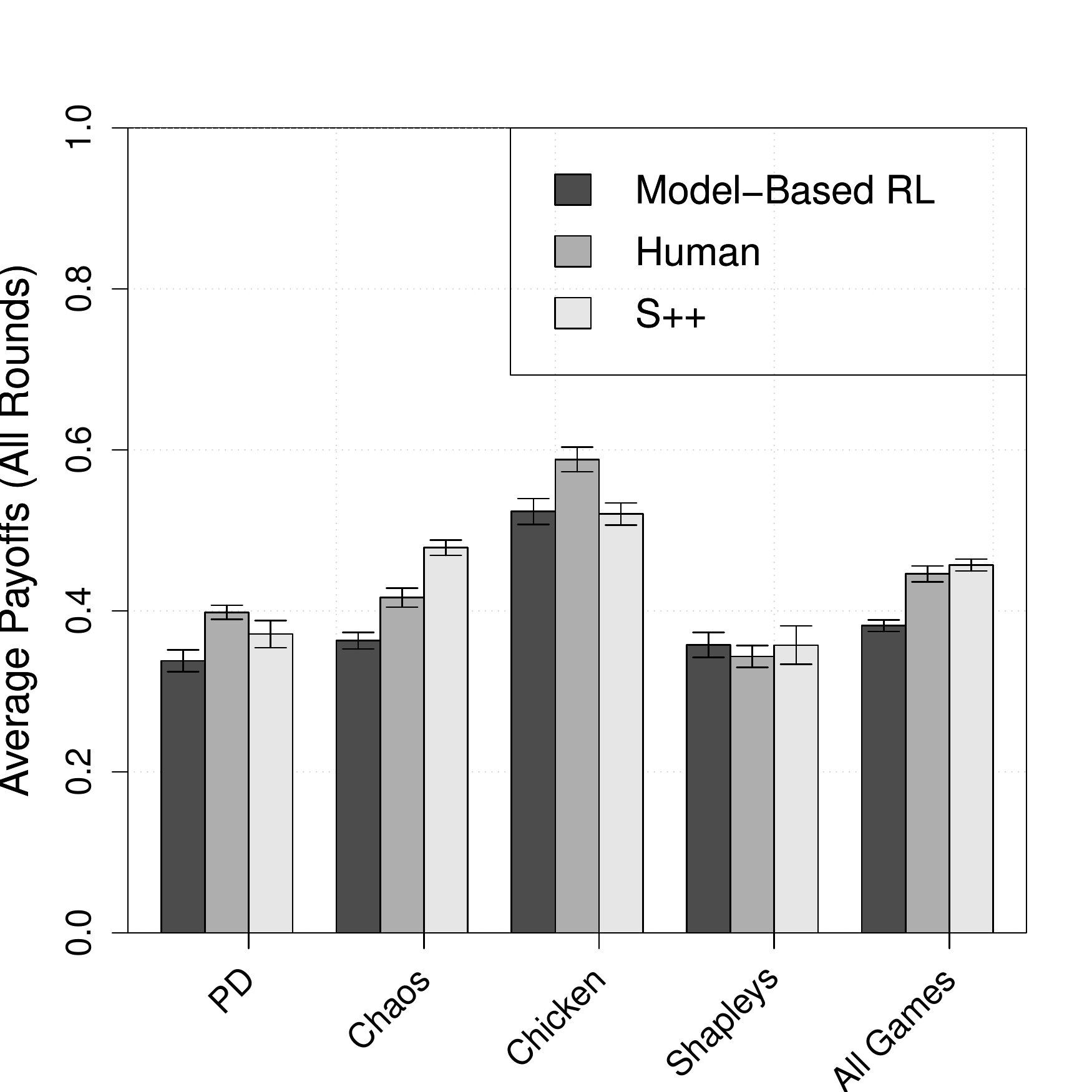}}~~~~
	\subfigure[vs.~S++: All rounds]{\label{fig:sBarChartAll}\includegraphics[width=2.25in]{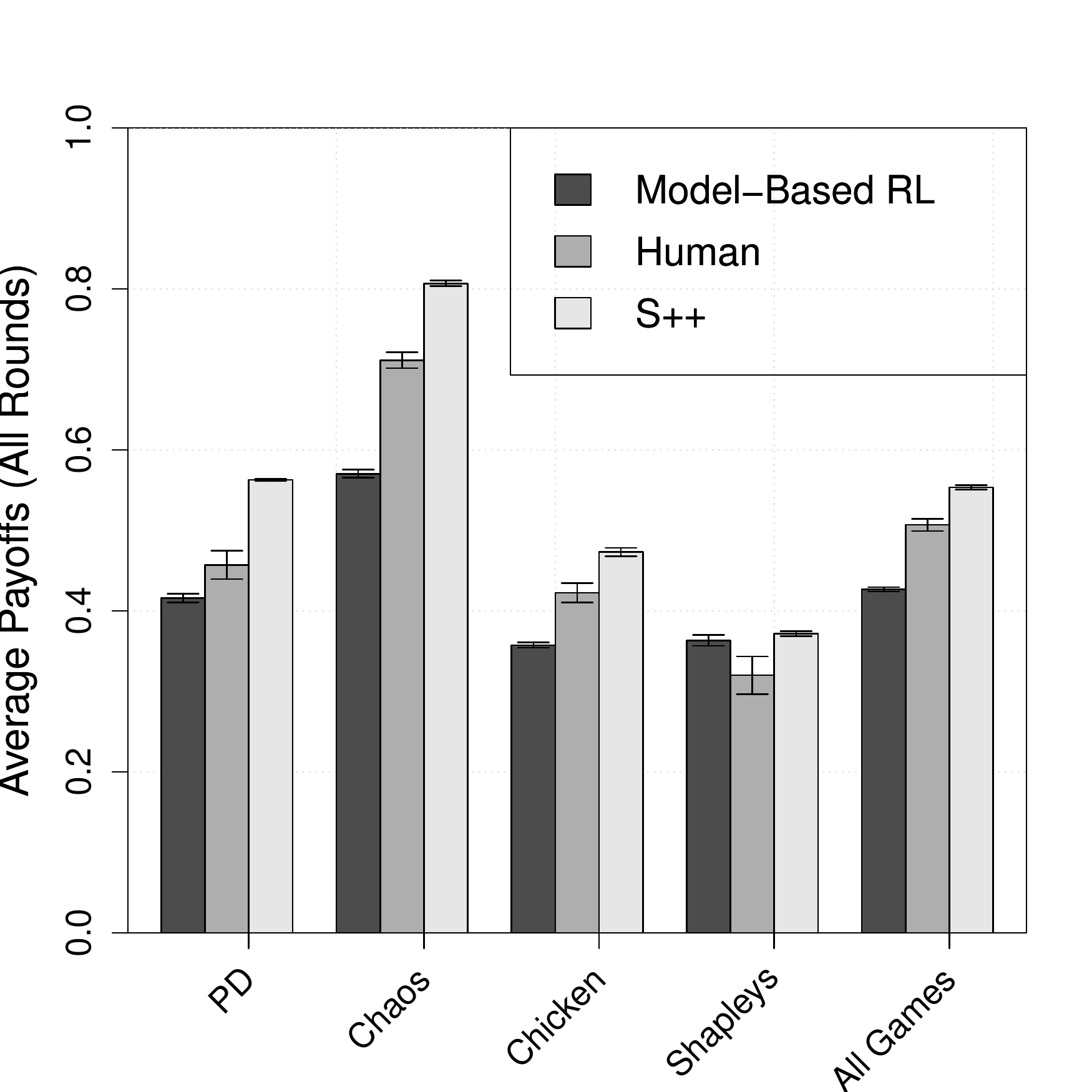}}~~~~
	\subfigure[vs.~Model-based RL: All rounds]{\label{fig:rBarChartAll}\includegraphics[width=2.25in]{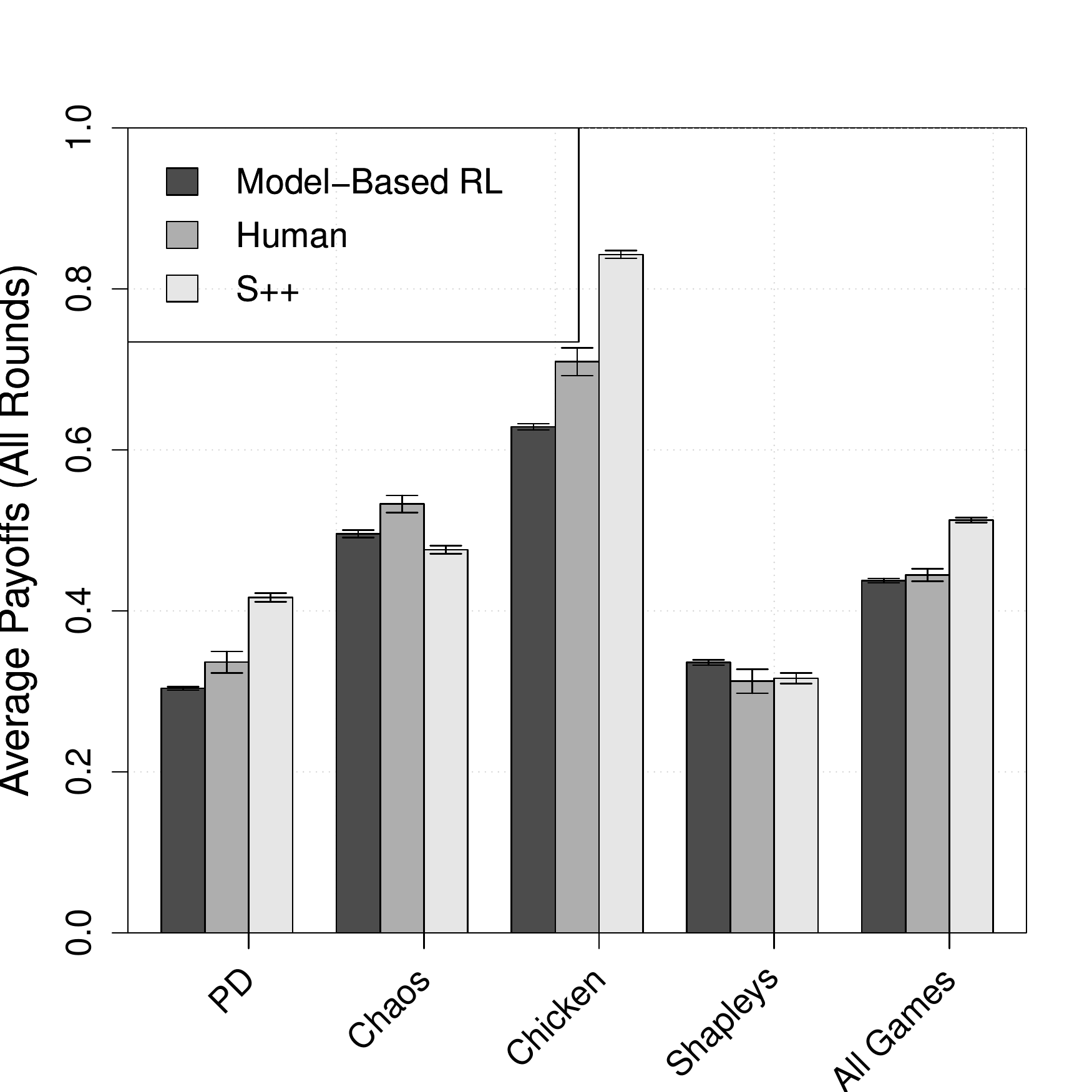}}\\ \vspace{-.2in}
	\subfigure[vs.~Humans: Last 10 rounds]{\label{fig:hBarChartLast}\includegraphics[width=2.25in]{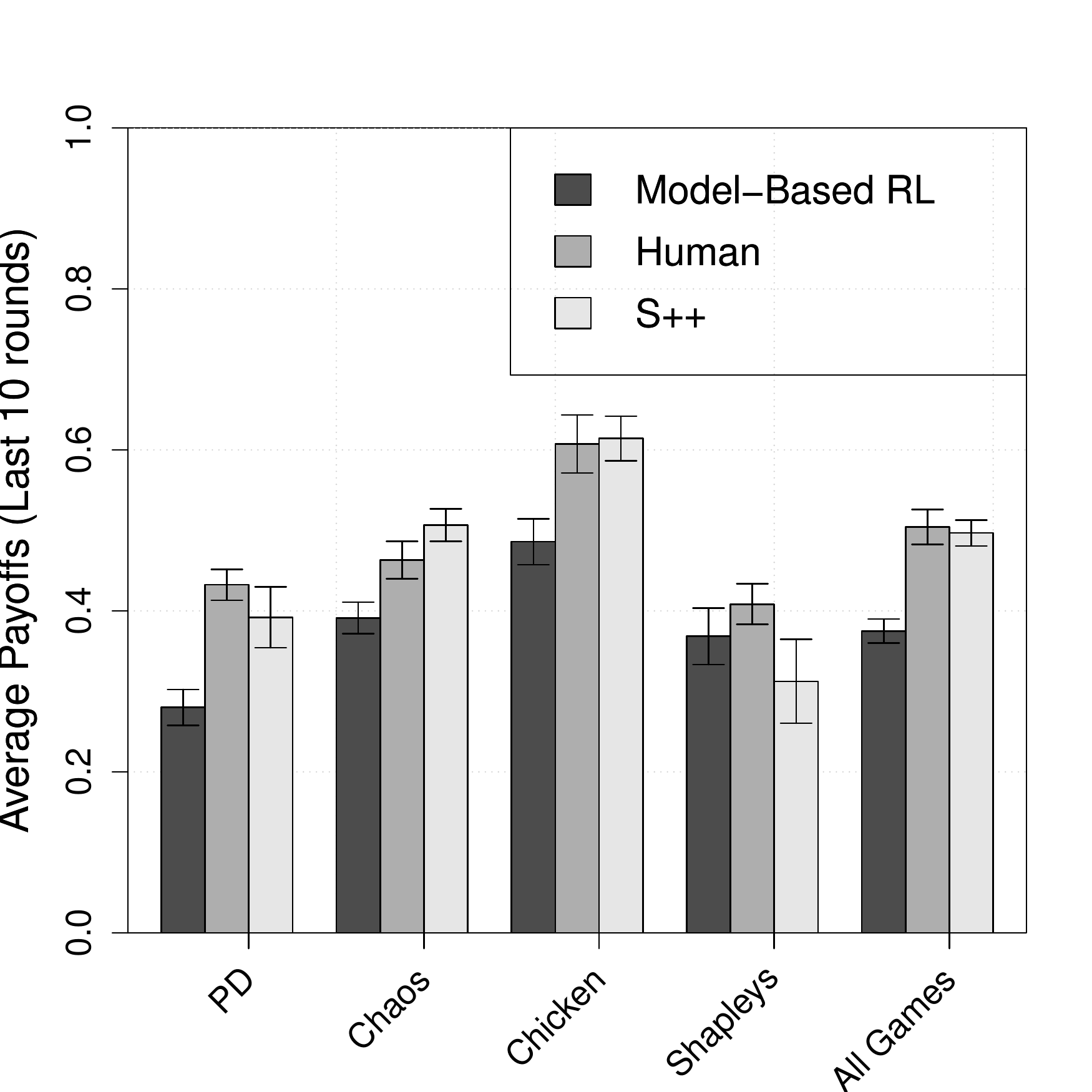}}~~~~~
	\subfigure[vs.~S++: Last 10 rounds]{\label{fig:sBarChartLast}\includegraphics[width=2.25in]{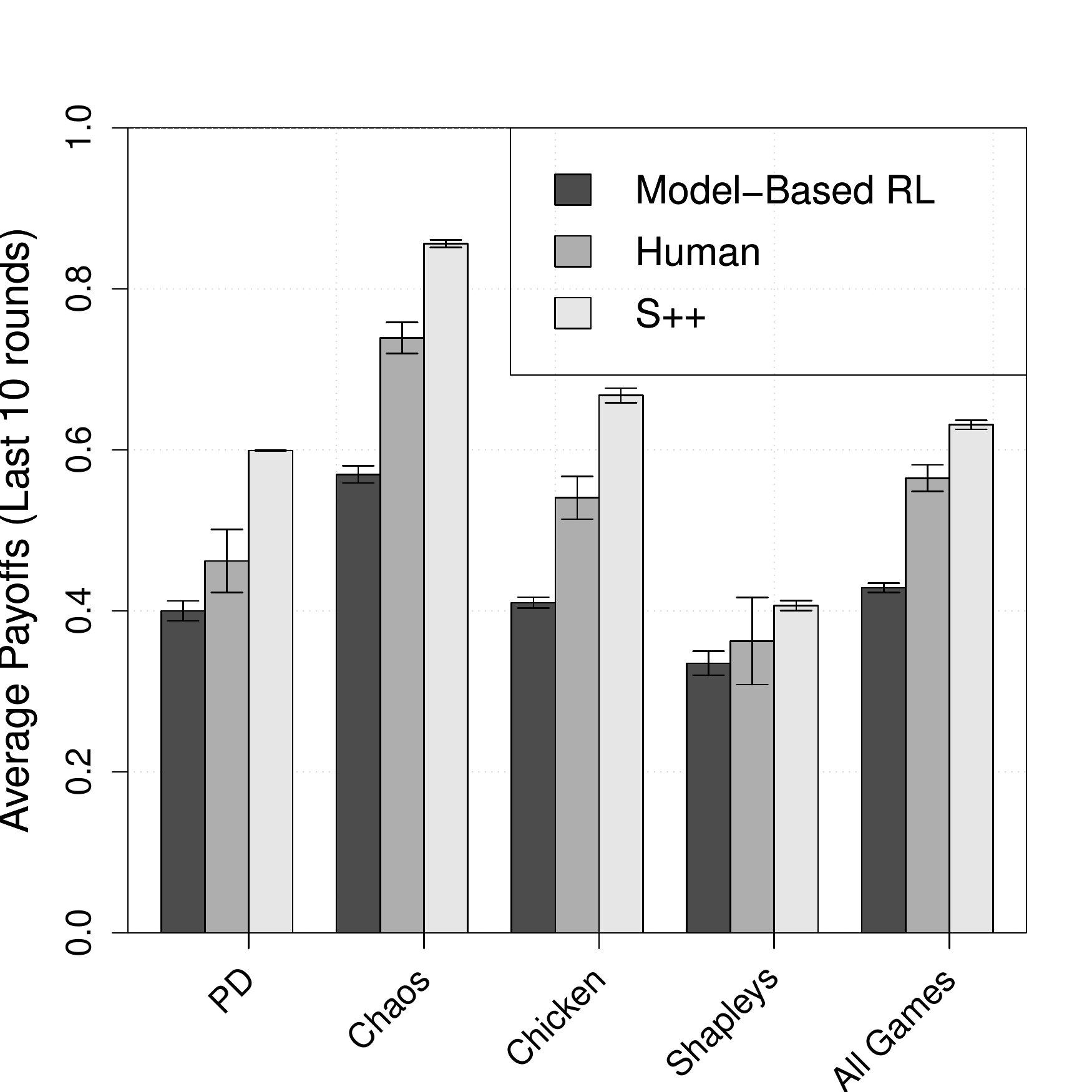}}~~~~~
	\subfigure[vs.~Model-based RL: Last 10 rounds]{\label{fig:rBarChartLast}\includegraphics[width=2.25in]{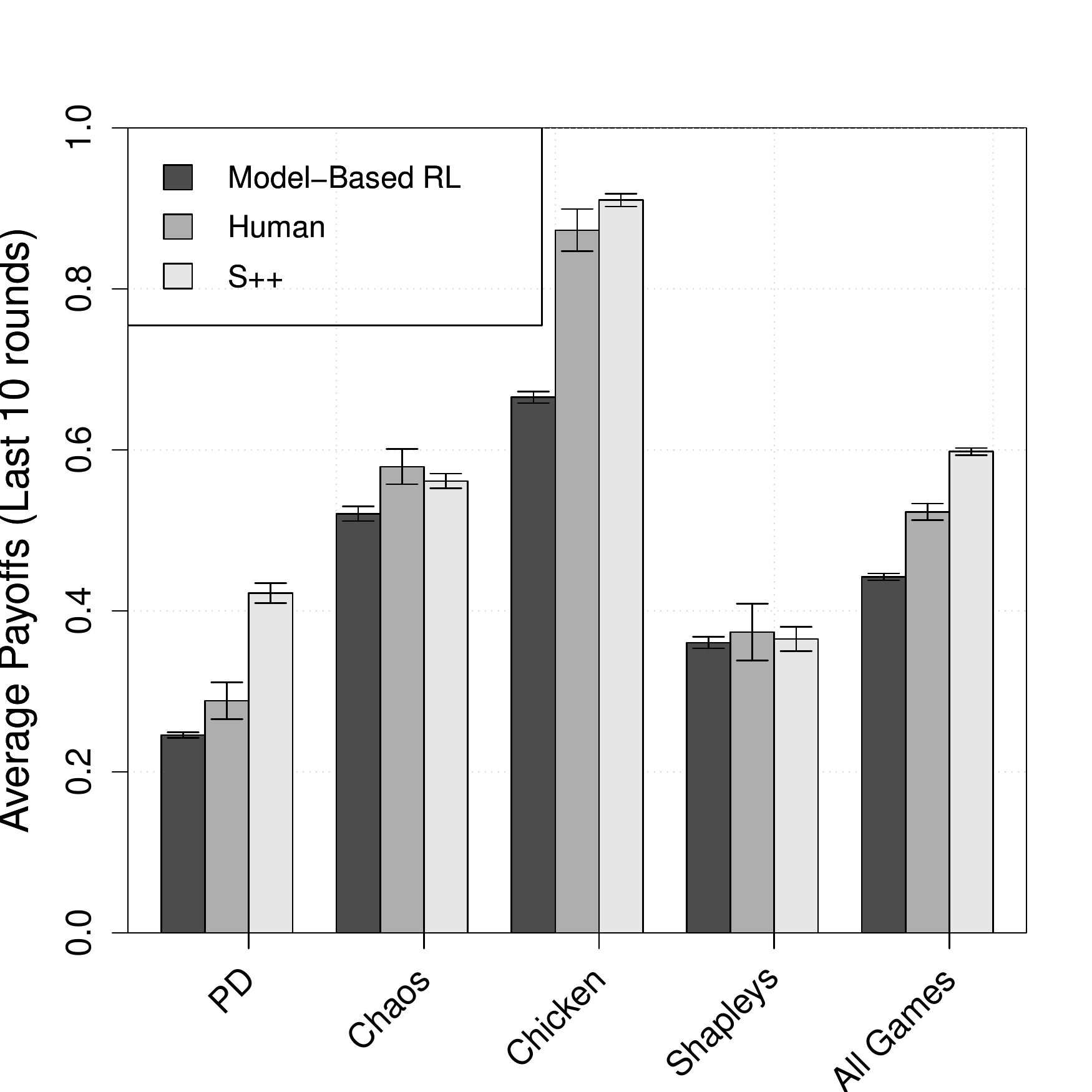}} \vspace{-.1in}
\caption{Mean payoffs for each pairing.  Error bars show standard error on the mean. In our experimental set-up, humans played as column players so all results of pairings vs.Humans represents those of the row player. While, the results for other players: S++ and Model-Based RL show the payoff to the row player. This distinction is especially important because of Chaos which is asymmetric.}
\label{fig:againstOthers}
\end{center}
\end{figure*}

The reason for the superiority of S++ over humans and Model-based RL in self play is indicated by Figure~\ref{fig:selfplay2}, which shows the average payoffs obtained by the agents as interactions progressed in each game.  The figure shows that, in each game, the average payoffs of S++ increased over time in self play.  These increases occurred as the S++ agents learned to cooperate and coordinate with each other.  On the other hand, the average payoffs of humans and Model-based RL did not consistently increase over time in these games.  People often failed to learn to make mutually beneficial compromises with each other in this study.

We do observe some variations by game.  While S++ clearly outperformed humans and model-based RL in self play in the PD and in Chaos over all rounds of the game, it was not as dominant in Chicken and Shapley's game over this time horizon.  For example, in Chicken, S++ spends the initial rounds of the game trying to bully its associate.  Once both agents find they cannot bully the other, they then begin to learn to play the solution $(a,a)$, which results in each player receiving a payoff of 0.83.  This is sometimes not achieved until after 50 rounds are completed.  Thus, in the long term, S++ performs very well in self play in this game, though it performs rather poorly in early rounds.

Coupled with the results shown in Figure 1, these results demonstrate the superiority of humans over older artificial learning processes (such as Model-based RL) in self play in tasks that require players to use intuitive judgment to establish effective collaboration in repeated interactions.  However, our results show that humans are not as effective in self play as S++, a recently developed AI algorithm for repeated interactions.  We believe that this represents a maturation in AI with respect to decision making in this domain.

\subsection{Against Each Other}
While performing well in self play is one important criterion of success, effective algorithms for repeated interactions must be able to learn to make profitable compromises when interacting with arbitrary associates.  The results of the pairings of the agents (humans, S++, and model-based RL) with each other help to establish how well each agent meets this objective.  The results of these pairings are summarized in Figure~\ref{fig:againstOthers}.  Against each kind of associate, we are interested in how the performances of S++ and Model-based RL compare to the performance of humans.

Figures~\ref{fig:hBarChartAll} and~\ref{fig:hBarChartLast} summarize the performance of the three agents when paired with humans.  These figures show that, overall, there was not a huge difference between the performance of humans and S++ against other humans.  Humans and S++ perform about the same overall.  However, Model-based RL had lower performance overall against humans than did S++ and humans.  Thus, S++ appears to interact with people in these games just as well as people interact with each other, while the older AI algorithm was not as effective.

When paired with another agent that is using S++, S++ outperformed both humans and Model-based RL on average in each of the four games (Figures~\ref{fig:sBarChartAll} and~\ref{fig:sBarChartLast}).  Additionally, consistent with the other scenarios in our study, people outperformed Model-based RL learning against S++.  The same overall trends are present when agents interacted with Model-based RL (Figures~\ref{fig:rBarChartAll} and~\ref{fig:rBarChartLast}), though results do vary from game to game.  S++ tended to outperform humans against Model-based RL.  Both S++ and humans had higher average payoffs overall against Model-based RL than Model-based RL achieves in self play.

In short, against all three types of associates, S++ either performed on par with or outperformed humans overall.  Meanwhile, both S++ and humans outperformed Model-based RL across these games.  These results speak to the maturation of AI algorithms.  At least in these games, our study indicates that new AI algorithms are beginning to learn to cooperate and compromise in repeated games at least as well as humans.


\section{Conclusions and Discussion}

New learning algorithms for playing general-sum repeated normal-form games are now able to converge in a reasonable number of rounds, thus enabling us for the first time to investigate their performance in interactions with people. We evaluated the agents and humans across the four normal-form games shown in Table~\ref{fig:gameMatrix}. These games (except for Chaos) are well-studied games from the literature, each of which represents a different common scenario. To be successful in each of these games, an algorithm must be able to learn to make and accept profitable compromises in many different situations.

Via a user study, we compared the performance of humans with Model-based RL and S++, a recently-published AI learning algorithm for repeated games.  In self play, S++ performed better than people and Model-based RL in all games except Chicken.  However, even in Chicken, S++ shows steady improvement in its performance as the game proceeds, so that its average payoffs eventually meet or eclipse those of humans and Model-based RL.  Furthermore, our study showed that S++ was as successful when paired with humans as humans are when paired with each other.  It also outperformed both humans and Model-based RL when paired with S++ and Model-based RL.  The consistency displayed by S++ shows an ability to learn to make the right choices in a variety of scenarios, and demonstrates that new AI algorithms appear to be gaining the ability to match humans in repeated interaction against arbitrary associates.

One of the weaknesses of our analysis is reliance on a small number of games. While this is an informative start, a comprehensive comparison between human and AI performance requires a more systematic investigation in a wider variety of games. Further, it is well-known that human performance in repeated games is highly sensitive to contextual factors, such as the framing of the game, the time given, or the presence of random noise \cite{rand2013human,Dal:2011,Fudenberg:etal:2012}. In the future, we plan to broaden our present study to investigate the role played by these factors in human-AI social interaction.

We note that AI algorithms were previously shown to compete effectively with humans in large zero-sum games such as Poker~\cite{PolarisPokerChampion}.  These games have unique equilibria characteristics that better support more traditional forms of game-theoretic reasoning.  It remains an open problem to create AI algorithms that can learn quickly enough to compete with humans in large repeated general-sum (stochastic) games played against arbitrary associates.

\section{Acknowledgement}
Manuel Cebrian is funded by the Australian Government as represented by the Department of Broadband, Communications and the Digital Economy and the Australian Research Council through the ICT Centre of Excellence program. We also thank Sohan DSouza for his help setting up the experimental platform.


\bibliographystyle{unsrt}
\bibliography{Bibliography}


\end{document}